\documentstyle[aps,prl]{revtex}
\begin{document}
\narrowtext
\twocolumn
\pagestyle{empty}

\noindent{\bf Comment on ``Quantum Pump for Spin and Charge Transport in 
a Luttinger Liquid''}

\bigskip

In their interesting paper \cite{1}, Sharma and Chamon address the  issue of
pumping
in the context of Luttinger liquids. They claim that for spinless
electrons  (i)  in the fast (UV)  limit, the charge pumped per cycle is
$0$ ($e$)  for repulsive, $g<1$,  (attractive, $g>1$) interaction; (ii)
this result is reversed in the adiabatic  (IR) limit.  Here we contest
(i).  We show  that this limit is non-universal, and that for any  given
pumping frequency $\omega_0$ and bare backscattering strength $\tilde X$
the  pumped charge is {\it larger} for the {\it smaller} $g$. We
complement our Comment by a  proof of  (ii).

Ref. \cite{1} considers a pump made of two gates placed a distance $2a$
apart and biased with ac voltages of the same frequency $\omega_0$. 
The pumping of spinless electrons is described 
\cite{1} by the low-energy limit of 
the bosonized action

\begin{eqnarray}
S=\int dt \Big\{\int dx \frac{v}{2g}\left [(\partial_x \Phi)^2+
\frac{1}{v^2}(\partial_t \Phi)^2\right ]+ & & \nonumber\\
X(t) \exp(i\sqrt{4\pi}\Phi(x=0))
+h.c. \Big\} & &
\end{eqnarray}
where the charge density is $e\partial_x \Phi/\sqrt{\pi}$. The charge
pumped per cycle is $q=e[\Phi(x=0,t=2\pi/\omega_0)-\Phi(x=0,t=0)
]/\sqrt{\pi}$.
$X(t)$ is an effective parameter that depends on 
the UV cutoff frequency $\Omega_c=v_F/a$
and the bare backscattering amplitudes
$\tilde X_1, \tilde X_2$ at the two constrictions. The renormalization group 
(RG) analysis of
Ref. \cite{2} shows that at the same bare backscattering 
amplitudes the renormalized
backscattering amplitude $X$ is greater for repulsive than 
for attractive electrons.

To estimate the charge $q$ pumped per cycle in the UV limit 
$\omega_0\sim\Omega_c$
we apply second order perturbation theory in $X$  \cite{2}. We find that 
$q\sim e [X/(hv)]^2 (\omega_0/\Omega_c)^{2g-2} \sim e [X/(hv)]^2$. This 
expression is valid as $X/(hv)\ll 1$. In this case $q=0$. 
As shown below at $X/(hv) \gg 1$ the charge pumped per cycle
is $e$. Thus, contrary to (i) the charge pumped per cycle is nonuniversal and 
can be either $0$ or $e$  (or any mean value between them)
for a given $g$. Moreover, the relation between $X$
and the bare backscattering amplitudes shows that for the same bare amplitudes
the pumped charge is larger the smaller is $g$.
This contradiction with (i) is due to
the peculiar definition of the UV limit employed in Ref. \cite{1}: 
$\omega_0\gg\omega_\Gamma$
where $\omega_\Gamma$ is a crossover frequency depending on $\Omega_c$ and $X$.
As $X$ is small (large) at $g>1$  ($g<1$) the crossover frequency
$\omega_\Gamma>\Omega_c$. Since $\omega_0<\Omega_c$, 
the UV limit in the sense of Ref.
\cite{1} does not exist in the above regimes. 
Thus, the UV limit employed in Ref. \cite{1}
is just the limit of strong (weak) backscattering at $g>1$ ($g<1$). It is
the backscattering amplitude and not the attractive or repulsive
nature of the interaction that
determines the pumped charge in the UV limit.

At the same time, (ii) is true and can be simply understood 
with the following argument.
We rewrite the backscattering contribution to the action as 
$\Gamma(t)\cos(\sqrt{4\pi}\Phi(0,t)-\alpha(t))$
where $\Gamma(t)>0$ is real. $\Gamma(t)$ and $\alpha(t)$ change at 
time intervals $\Delta t\sim 1/\omega_0$.
At shorter time-scales the RG approach of 
Ref. \cite{2} can be used since $\Gamma$
and $\alpha$ may be considered as time-independent.
One finds that $\Gamma$ grows (decreases) under the 
action of the RG at $g<1$ ($g>1$), while the phase $\alpha$ remains unchanged.
At $g>1$ at the RG scale $\omega_0$ we 
obtain a renormalized action with a small $\Gamma_R(t)$.
The Keldysh perturbative expansion in $\Gamma_R(t)$ \cite{3}
shows that the charge pumped per cycle is much smaller than $e$.
At $g<1$ we obtain a large renormalized $\Gamma_R(t)$ at the scale $\omega_0$. 
Hence, fluctuations about the minimum
of the backscattering term in the action 
$(\sqrt{4\pi}\Phi(x=0)-\alpha)=-\pi/2$ are small. The charge
pumped per cycle is $q=e\Delta\Phi(0)/\sqrt{\pi}=e\Delta\alpha/(2\pi)=en$, 
where $\Delta\alpha=2\pi n$
is the change of the phase $\alpha$ per period.

In conclusion we confirm the result (ii) 
for the IR limit but show that the prediction
(i) about the UV case as well as related predictions for electrons with spin
are invalid and due to a peculiar definition 
of the UV limit used in Ref. \cite{1}. 

This work is supported by 
GIF foundation, the US-Israel Bilateral Foundation,
the ISF of the Israel Academy (Center of Excellence), and by the DIP Foundation.
DEF acknowledges the support by 
the Koshland scholar award and RFBR grant 00-02-17763.

\bigskip
\noindent Dima E. Feldman$^{1,2}$ and Yuval Gefen$^1$ 
\newline {$^1$\small Condensed Matter Physics Department,
Weizmann Institute of Science, 76100 Rehovot, Israel}
\newline{$^2$\small Landau Institute for Theoretical Physics, Chernogolovka} 
\newline{\small 142432 Moscow region, Russia}

\bigskip
\noindent Received xxx \newline \noindent PACS numbers {71.10.Pm,
72.25.-b, 73.63.Fg, 73.63.Nm }


\begin{references}
\frenchspacing
\bibitem{1}
P. Sharma and C. Chamon, Phys. Rev. Lett. {\bf 87}, 096401 (2001).
\bibitem{2}
C.L. Kane and M.P.A. Fisher, Phys. Rev. B {\bf 46}, 15233 (1992).
\bibitem{3}
D.E. Feldman and Y. Gefen, e-print cond-mat/0104558.

\end{references}
\end{document}